# Gas sensing potential of stacked graphene/h-BN structures: a DFT-based investigation

Martin Siebel, Pavel Rubin*, Raivo Jaaniso

Institute of Physics, University of Tartu, W. Ostwaldi 1, 50411 Tartu, Estonia

**Abstract**

Using periodic DFT, we examined the adsorption of $NO_2$, $NH_3$, and $O_3$ on the h-BN side of a graphene/h-BN heterostructure designed as a model gas-sensor material. The h-BN overlayer serves both as an active adsorption surface and as protection that may reduce irreversible processes such as graphene oxidation. Two model systems were considered: an extended graphene/h-BN bilayer ($B_{36}N_{36}C_{72}$) and a graphene sheet partially covered by a smaller h-BN island ($B_{11}N_{11}C_{72}$). Their electronic structures differ strongly near the Dirac point. In the extended bilayer, the Fermi level remains aligned with that of pristine graphene, indicating negligible charge transfer. In the island-covered system, the Fermi level shifts to lower energies, reflecting electron transfer from graphene to h-BN. These differences lead to distinct adsorption behavior. $NO_2$ binds much more strongly to $B_{11}N_{11}C_{72}$, forming a chemical bond, while $O_3$ dissociates on this surface but remains intact on the extended bilayer. $NH_3$ unusually acts as an electron acceptor in the island system. Overall, $NO_2$ and $O_3$ substantially increase graphene conductivity, whereas $NH_3$ induces much weaker changes. These results highlight the potential of graphene/h-BN heterostructures for gas sensing.

* Corresponding author
  *E-mail address:* rubin@ut.ee

# 1. Introduction

Highly sensitive and selective gas sensors are needed across a range of applications, including environmental protection, industrial process control, and medical diagnostics. Graphene and related two-dimensional (2D) materials are promising building blocks for next-generation gas sensors due to their high surface-to-volume ratio and outstanding electronic properties [1]. The growing diversity of 2D materials, along with the feasibility of their large-area synthesis [2] and incorporation into semiconductor production lines [3], is advancing rapidly. Heterostructures of materials with thicknesses of one or a few atomic layers enable the separation of the sensor's subfunctions: one material serves as a selective receptor for gas molecules, and another as a highly sensitive electronic transducer [4]. In environments containing aggressive gases, the stability of the latter must be ensured, which can be provided by an inert 2D over- or interlayer.

Graphene, owing to its low charge-carrier density yet high electrical conductivity, is an exceptionally sensitive electronic transducer for chemical sensors, affording a high signal-to-noise ratio [5]. However, graphene is prone to oxidation, especially at elevated temperatures and in the presence of reactive oxygen species [6]. By contrast, the so-called "white graphene"—hexagonal boron nitride (h-BN)—is much more resistant [7]. In this work, we investigate structures that exploit graphene's ultra-high sensitivity to charge rearrangements or to changes in local electric fields occurring in its vicinity, while simultaneously being resistant to oxidation thanks to the properties of an h-BN overlayer on graphene.

Following the experimental demonstration of detecting single gas-molecule adsorption on graphene [5], numerous experimental and theoretical studies have focused on modifying graphene to enhance its gas-sensing performance. In particular, Zhang et al. [8] conducted an *ab initio* computational study to investigate the effects of defects, such as vacancies, and dopants on graphene. Their first-principles calculations demonstrated that these modifications enhance gas adsorption by creating new active sites for molecule binding and alter the electronic structure and charge distribution, leading to more pronounced conductivity changes upon gas adsorption. Consequently, the sensitivity of graphene-based sensors is significantly improved, underscoring the potential of defect- and dopant-engineered graphene for gas detection applications. Consistent with these results, numerous other computational studies have reported similar conclusions, notably in references [9,10]. In turn, a number of experimental investigations have been reported in this area. In particular, in [11], it was demonstrated that depositing thin nanostructured Au and Pt layers on epitaxial graphene enables tuning of its electronic properties and significantly enhances gas sensitivity (particularly to $NO_2$) through Fermi level ($E_F$) adjustment and the creation of additional active adsorption sites. Srivastava et al. [12] demonstrated that boron-doped few-layer graphene, synthesized by low-pressure chemical vapor deposition and characterized by spectroscopic and microscopic methods, exhibits rapid and reversible ammonia sensing at room temperature. Functionalization of graphene with a few layers of carefully selected metal oxides, acting as receptors and redox catalysts, has demonstrated excellent sensing performance [13,14].

Our work is focused on the theoretical investigation of the sensory properties of a distinct class of materials: graphene/hexagonal boron nitride heterostructures [15-17]. These materials exhibit a layered structure analogous to graphene, consisting of hexagonally arranged planes held together by van der Waals interactions [15]. Within each h-BN plane, every boron atom is covalently bonded to three neighboring nitrogen atoms. The lattice structures of graphene and h-BN are closely matched, with a lattice mismatch of approximately 1.5%. The interest in graphene/h-BN heterostructures arises from several factors. Hexagonal boron nitride possesses a wide bandgap of 5.9 eV [15] and functions as a dielectric, making it an ideal substrate for graphene. Moreover, h-BN can modify the electronic and optical properties of graphene; under specific conditions, it can induce a finite bandgap in graphene, which is crucial for nanoelectronic applications. In addition, h-BN protects graphene from degradation by reactive oxygen species, particularly ozone. Experimental studies have demonstrated the feasibility of fabricating both stacked and in-plane graphene/h-BN heterostructures [15-19]. In particular, graphene/h-BN stacked heterostructures were fabricated either by transferring mechanically exfoliated graphene onto exfoliated h-BN single crystals deposited on SiO$_2$ substrates, or by depositing graphene grown via chemical vapor deposition (CVD) onto exfoliated commercial h-BN (Momentive AC6004) [18]. In addition, hexagonal boron nitride films used for graphene/h-BN in-plane heterostructures can be synthesized by CVD on Cu or Ni foils using ammonia borane as a precursor, producing high-quality layers suitable for device fabrication [19].

Several studies have reported theoretical calculations of the electronic structure of these systems. In-plane heterostructures can exhibit a wide range of geometries, including graphene superlattices with BN quantum dots [20], carbon islands and nanoribbons embedded within BN sheets [21], as well as graphene/h-BN heterointerfaces [22, 23], among other configurations. Density functional theory (DFT) calculations in [20] reveal the presence of a nonzero band gap in BN-embedded graphene, independent of the edge structure, superlattice symmetry, or BN quantum dot configuration. De Souza *et al.* [22] analyzed single layers of graphene and h-BN, as well as in-plane fused stripes, in contact with NO, NO$_2$, NH$_3$, and CO$_2$ molecules by positioning the adsorbates at specific surface sites. Among the studied species, NO$_2$ exhibited the strongest interaction, with a binding energy of up to 2 eV and an equilibrium distance of 1.53 Å, accompanied by a measurable change in conductance. In contrast, the adsorption of NH$_3$ and CO$_2$ was significantly weaker, with binding energies of approximately 0.33 eV and equilibrium distances of 2.86 Å and 2.85 Å, respectively. Similarly, Mawwa et al. [23] investigated pristine graphene, h-BN, and their in-plane heterostructures composed of eight hexagons with edges saturated by 14 hydrogen atoms. For CO and SO$_2$ adsorption, the interaction energies were found to range from 0.17 to 0.41 eV, with equilibrium gas–substrate separations of 3.05–3.37 Å. Interestingly, the binding energy was substantially larger for CO, whereas the adsorption distance was longer than for SO$_2$.

The majority of ab initio studies on stacked graphene/h-BN heterostructures [24–29] have been conducted under the simplifying assumption that graphene and h-BN possess identical lattice constants. When graphene and h-BN planes are arranged such that one carbon atom lies above a boron site and the other above the center of a BN hexagon—an energetically favorable configuration—a bandgap of 53 meV opens in the graphene electronic spectrum [24]. In this configuration, the Fermi level is located within the induced bandgap. This gap originates from the inequivalence of the two carbon sublattices within the considered structure, which consists of four h-BN layers capped by a graphene sheet. The equilibrium interlayer spacing between graphene and h-BN is 3.22 Å. As demonstrated in [25] for analogous structures, the magnitude of the induced gap is strongly dependent on the interlayer separation between the graphene and h-BN subsystems and stacking arrangement. In [26], density functional theory (DFT) calculations employing the plane-wave pseudopotential method were used to investigate the electronic structure of a graphene monolayer supported on a single layer of h-BN. The results revealed that this substrate induces a small band gap of approximately 57 meV in graphene. Furthermore, sizeable band gaps at the Dirac point have also been reported for stacked graphene/h-BN heterostructures in [27, 28]. Extensive first-principles calculations have been carried out by Zollner, et al. [29] for several stacking configurations of Gr on h-BN, and for h-BN encapsulating Gr, without addressing the sensor related properties. DFT calculations were done also for incommensurate systems [30]. Results demonstrate essential diminishing of gaps in the graphene subsystem in comparison to the commensurate case. The sensing properties of graphene/h-BN heterostructures toward CO, CO$_2$, NO, and NO$_2$ were investigated theoretically in [31]. The study reported binding energies of NO$_x$ and CO$_x$ molecules on graphene supported by h-BN in the range of 0.2–0.3 eV. As anticipated, the h-BN substrate exerted only a minor influence on the physisorption energies.

In this work, we theoretically investigate the feasibility of h-BN/graphene heterostructures for gas-sensing applications, focusing on a configuration in which the h-BN layer is positioned above the graphene plane. Graphene/h-BN heterostructures are well known for their use in high-mobility field-effect transistors, where hexagonal boron nitride encapsulates graphene from both the top and bottom, providing atomically flat dielectric interfaces that reduce carrier scattering and enhance charge carrier mobility. Alternatively (as in [31]), in a graphene-on-h-BN configuration, with h-BN as the bottom substrate and graphene exposed on top, the structure is employed in gas sensors, where the clean h-BN support preserves graphene's electronic quality while its exposed surface enables highly sensitive detection of gas molecules through changes in conductivity. In our approach, the h-BN layer, which hosts the adsorbed molecules, acts as a chemically inert protective barrier that mitigates graphene–atmosphere interactions. As a result, the intrinsic electronic properties of graphene are preserved, irreversible degradation pathways are suppressed, and long-term stability for electronic applications is enhanced. Density functional theory (DFT) calculations were performed for planar "double-decker" structures composed of two-dimensional graphene and h-BN, namely $C_{72}B_{36}N_{36}$ and $C_{72}B_{11}N_{11}$, as well as their interactions with the small molecules $NO_2$, $NH_3$, and $O_3$. In chemiresistive toxic gas sensors, graphene functions as the conductive channel, whereas reactive gas molecules interact directly with the dielectric h-BN layer, which exhibits adsorption properties distinct from those of graphene. The sensor signal arises from changes in the electron concentration and mobility within graphene, induced by interactions between gas molecules and the h-BN/graphene heterostructure.

## 2. Computational method

The adsorption of $NO_2$, $O_3$, and $NH_3$ molecules on the h-BN/graphene sensor was investigated by means of quasi-two-dimensional density functional theory (DFT) non spin-polarized calculations employing periodic boundary conditions. All calculations were performed using the Vienna Ab initio Simulation Package (VASP) [32, 33] with the projector augmented-wave (PAW) method [34, 35]. The exchange–correlation interactions were described within the PBEsol functional [36], and long-range van der Waals interactions were accounted for using the DFT-D3 correction scheme [37]. In modeling the system, two representative structures were considered. The first model consisted of two infinitely extended graphene and h-BN planes, represented by a $B_{36}N_{36}C_{72}$ periodic cluster. In this case, the small lattice mismatch (<2%) between graphene and h-BN was neglected. The second model comprised an infinite graphene layer with a smaller h-BN layer on top, described as $B_{11}N_{11}C_{72}$. The supercell dimension perpendicular to the planes was fixed at 30 Å. During geometry optimization, all atoms except carbon were allowed to fully relax, while carbon atoms were constrained along the c-direction. The supercell lattice parameters (constant volume and shape) and atomic coordinates were optimized until the total energy difference between consecutive ionic steps was less than $5.0 \times 10^{-5}$ eV. The resulting equilibrium structures exhibited average residual forces below 0.02 eV/Å per atom. Ground-state structures of these clusters were obtained using periodic boundary conditions. A plane-wave cutoff energy of 500 eV was employed together with a Γ-centered mesh for k-point sampling. Structural relaxations and stability analyses were performed with a $4 \times 4 \times 1$ k-point grid, whereas the electronic structure, specifically the density of states, was evaluated on a denser $25 \times 25 \times 1$ grid to achieve higher accuracy. Subsequently, adsorption was studied by introducing the molecules (M = $NO_2$, $NH_3$, $O_3$) onto both cluster models to form $M@B_{36}N_{36}C_{72}$ and $M@B_{11}N_{11}C_{72}$ complexes. The adsorption energy ($E_{ads}$) was determined according to:

$$E_{ads} = E_{complex} - (E_{Gr/h\text{-}BN} + E_M), \tag{1}$$

where $E_{complex}$, $E_{Gr/h\text{-}BN}$ and $E_M$ represent the total energies of the adsorbed system, pristine substrate, and isolated gas molecule, respectively. The Bader charge analysis method [38] was employed to evaluate charge redistribution in the system upon gas adsorption. Examination of the charge transfer, together with changes in the partial density of states (PDOS), provides insight into the sensitivity of the graphene/h-BN system to the adsorbed molecules.

## 3. Results and Discussion

**DFT simulation of $B_{36}N_{36}C_{72}$ and $B_{11}N_{11}C_{72}$ structures.**

The stacked structures $B_{36}N_{36}C_{72}$ and $B_{11}N_{11}C_{72}$, which model the gas sensor, are shown in Fig. 1(a) and Fig. 1(b). In the $B_{36}N_{36}C_{72}$ system, the stacking configuration places the B and N atoms directly above the carbon atoms. After full relaxation, the interlayer distance between BN and graphene is found to be 3.46 Å, which is in good agreement with previous studies on graphene/h-BN heterostructures, particularly the value of 3.5 Å reported in [24]. The atomic arrangement of the structure with a BN layer fragment is shown in the inset of Figure 1(b). The average separation between the BN fragment and graphene is approximately 3.28 Å, while the shortest distance between atoms of the BN fragment and the graphene plane is 3.08 Å. Variations in the spatial configuration of the systems lead to corresponding changes in their electronic structure. Bader charge analysis (Table 1) indicates the absence of charge transfer between the carbon and h-BN subsystems in the $B_{36}N_{36}C_{72}$ system. This conclusion is further supported by the partial density of states (PDOS) shown in Figure 2(a), where the Fermi level of the heterostructure aligns with the Dirac point of the carbon subsystem, in close analogy to pristine graphene. In the top BN layer, nearly complete charge transfer of valence electrons occurs from boron to nitrogen. This result is consistent with charge density studies of h-BN obtained from synchrotron radiation powder diffraction using the maximum entropy method, which yielded ionic charges of +2.7 e for boron and −1.9 e for nitrogen [39]. In analogy with [24], the presence of the h-BN plane induces a small gap at the Dirac point of approximately 70 meV. In the $B_{36}N_{36}C_{72}$ system the HOMO–LUMO separation of the h-BN subsystem is about 4.5 eV, which is comparable to the band gap values reported for bulk (3D) h-BN [40]. As shown in Table 1, the $B_{11}N_{11}C_{72}$ cluster exhibits transfer of 0.28e from the carbon plane to the BN fragment. Unlike the complete structure with two infinite planes, the upper BN fragment in this case has a finite size, resulting in boundaries with unsaturated nitrogen and boron atoms. Consequently, multiple peaks with different energies arise, corresponding to boron- and nitrogen-related states. This change in behavior can be attributed to the increased localization of electrons, in contrast to the delocalized electronic distribution observed in the infinitely extended structure. The HOMO–LUMO separation of the BN subsystem is ~1.5 eV. Unlike the h-BN layer, the graphene sheet remains infinite, and its PDOS is mainly consistent with that of the $B_{36}N_{36}C_{72}$ system. As shown in Fig. 2(b), the Fermi level shifts by ~0.4 eV below the Dirac point, reflecting charge redistribution between subsystems. As shown in Figure 2(b), the V-shaped PDOS of graphene near the Fermi level, characteristic of pristine graphene, is distorted. At the same energy, pronounced PDOS peaks of Ni and B are observed. Together with the charge transfer, this indicates hybridization between the electronic states of graphene and h-BN. The small magnitude of charge transfer in both considered structures suggests that stabilization is primarily governed by van der Waals interactions described by the DFT-D3 correction. Thus, the systems are suitable models for gas sensing, and their responses toward $NO_2$, $NH_3$, and $O_3$ are discussed below.

Table 1 Charge distribution in the $B_{36}N_{36}C_{72}$ and $B_{11}N_{11}C_{72}$ systems. Bader analysis. The "Initial" column shows the number of valence electrons included in the PAW pseudopotentials.

| Element | $B_{36}N_{36}C_{72}$ | | $B_{11}N_{11}C_{72}$ | |
|---|---|---|---|---|
| | Initial | Bader charge analysis | Initial | Bader charge analysis |
| C | 288 | 288.0113 | 288 | 287.7272 |
| B | 108 | 0.0007 | 33 | 0.0005 |
| N | 180 | 287.9879 | 55 | 88.27228 |

**$NO_2$ adsorption.**

The adsorption energies of a single $NO_2$ molecule on $B_{36}N_{36}C_{72}$ and $B_{11}N_{11}C_{72}$ are −1.0290 and −3.8469 eV (Fig. 3 (a) and Tables VII and VIII in the Supplementary Material, Tables I–VIII are provided in the Supplementary

Material). The corresponding complexes are shown in Fig. 1(c) and Fig. 1(d), respectively. In the $NO_2@B_{36}N_{36}C_{72}$ complex, the shortest B–N($NO_2$) and N–O distances are 3.40 and 3.26 Å, whereas in $NO_2@B_{11}N_{11}C_{72}$ these distances decrease to 1.58 and 1.47 Å. These results indicate that $NO_2@B_{36}N_{36}C_{72}$ corresponds to physical adsorption, while the other structure represents chemical adsorption. The sensor response to $NO_2$ adsorption, expressed as a change in conductivity, arises from modifications in the electronic structure of the graphene subsystem. As shown in the bar graphs in Fig. 3 (b - d) (see also more detailed data in the Supplemental Material), adsorption in both structures induces electron transfer from the graphene layer: –0.2193e for $NO_2@B_{36}N_{36}C_{72}$ and –0.1258e for $NO_2@B_{11}N_{11}C_{72}$. The stronger charge transfer in the physically adsorbed case is accompanied by a more pronounced Fermi-level shift and a larger increase in the PDOS of the graphene subsystem at the Fermi level (Fig. 2(c)). According to the Kubo–Greenwood formalism [41], this leads to enhanced conductivity, with the effect being more significant for physical adsorption. In the $NO_2@B_{36}N_{36}C_{72}$ structure, adsorption does not affect the charge of the intermediate BN layer; instead, charge is transferred from the graphene layer to the adsorbed molecule. In this case, the Fermi level remains close to the Dirac point, and the tunneling transfer underlying this redistribution can be explained by the Klein paradox, a characteristic feature of graphene [42]. Accordingly, the V-shaped density of states of the graphene subsystem remains essentially unchanged upon adsorption. In contrast, in the $NO_2@B_{11}N_{11}C_{72}$ structure, charge transfer involves both the graphene and BN layers (PDOS of this complex is shown in Fig. 2(d)). The charge transfer from the graphene plane, and the corresponding change in conductivity, are significantly larger in the ideal bilayer structure compared to $NO_2@B_{11}N_{11}C_{72}$. It is reasonable to compare our results with DFT calculations of $NO_2$ adsorption on pristine graphene and in-plane graphene/BN heterostructures. First-principles studies consistently show that $NO_2$ acts as an electron acceptor when adsorbed on graphene [8, 43]. In particular, Ref. [43] reports physisorption with an adsorption energy of approximately −0.07 eV and a charge transfer of about 0.1e from graphene to the $NO_2$ molecule, resulting in p-type doping of graphene and a corresponding shift of the Fermi level toward lower energies. Reference [8] predicts a larger adsorption energy (−0.48 eV) and a higher charge transfer (−0.19e), which are closer to our result for infinite BN pane. Also these comparisons indicate that adsorption on the BN layer leads to enhanced interaction strength, since $NO_2$ interacts with a polar surface composed of partially charged B and N atoms, resulting in a stronger electrostatic contribution than in the case of pristine graphene. In G/h-BN /G/h-BN/G in-plane heterostructures [44], DFT simulations show that gas molecules preferentially adsorb on B sites of the h-BN regions. The adsorption energies of $NO_2$ molecules are reported to be around 1.1 eV, with a charge transfer of approximately 0.3e from the surface to the molecule, which is close to our result. A similar model was considered in Ref. [22], where even larger adsorption energies were obtained (about -2 eV), accompanied by a substantial charge transfer of approximately 0.79 |e|. Overall, the results reported for pristine graphene and in-plane heterostructures are qualitatively consistent with our findings. However, our stacked model offers an important advantage in that it protects the graphene layer from degradation while preserving strong adsorption characteristics.

**$O_3$ adsorption.**

Figures 1(e) and 1(f) present the $O_3@B_{36}N_{36}C_{72}$ and $O_3@B_{11}N_{11}C_{72}$ complexes. Ozone adsorption on $B_{36}N_{36}C_{72}$ closely follows the behavior of $NO_2$, with a charge transfer of –0.2456 e from graphene (the bar graphs in Fig. 3(b, c) and Table III) occurring directly between graphene and oxygen, bypassing the BN plane. The minimum N–O and B–O distances are 2.96 Å and 3.08 Å, respectively, and the adsorption energy is –0.3759 eV (the bar graph in Fig. 3(a) and Table VII) indicating physical adsorption. In contrast, adsorption on $B_{11}N_{11}C_{72}$ leads to ozone dissociation into a charged $O_2$ complex (B–O 1.52 Å) and an oxygen ion (B–O 1.28 Å), which acquire additional charges of 0.829 e and 1.724 e, respectively, while graphene loses –0.5064 e (the bar graph in Fig. 3(b, d) and Table IV). The oxygen subsystem accepts charge from both planes, and the large overall adsorption energy of –5.505 eV (Table VII) confirms chemical adsorption. Partial DOS (Figures 2(e) and 2(f)) reveal a downward shift of the Fermi level and increased graphene PDOS in both cases, demonstrating that these structures are capable of detecting ozone molecules. It is worth noting that Ref. [45] demonstrates that dissociation is already possible in the case of $O_3$ adsorption on pristine graphene. In that work, ozone ($O_3$) was shown to physisorb on graphene with a binding energy of approximately −0.46 eV when van der Waals interactions were included. After overcoming an activation barrier of about 0.75 eV, $O_3$ can chemisorb by forming an epoxy group on the graphene lattice while simultaneously releasing an $O_2$ molecule. This process results in a more stable chemisorbed state with an adsorption energy of approximately −0.60 eV and leads to p-type doping of graphene. Density functional theory (DFT) simulations of doped BN also demonstrate the possibility of ozone dissociation [46]. On carbon-doped BN, ozone adsorbs weakly ($E_a \approx -0.27$ eV) without dissociation, whereas on silicon-doped BN it dissociates, with one oxygen atom chemisorbing on the surface while an $O_2$ molecule is released. This process results in a strong adsorption energy of approximately −8.07 eV. Phosphorus doping similarly promotes ozone dissociation accompanied by $O_2$ release. These findings indicate that dopant-induced surface reactivity can facilitate ozone dissociation and lead to strong chemisorption. Notably, this large adsorption energy is comparable to our result for the $O_3@B_{11}N_{11}C_{72}$ system, where the charge transferred from the BN fragment to the oxygen subsystem (−2.048 |e|) is significantly greater than the charge transferred from graphene. In contrast, system $O_3@B_{36}N_{36}C_{72}$, featuring a pristine BN plane, exhibits chemical inertness, and the charge transfer occurs primarily from the graphene layer. Upon relaxation, the $B_{11}N_{11}$ fragment deviates from the ideal planar structure, which increases its surface reactivity and creates favorable conditions for ozone dissociation. Thus, the comparison of systems $O_3@B_{36}N_{36}C$ and $O_3@B_{11}N_{11}C_{72}$ highlights that structural deformation plays a key role in enabling $O_3$ dissociation and strong chemisorption in graphene/BN heterostructures.

**$NH_3$ adsorption.**

As in the previous cases, $NH_3$ adsorption differs significantly between the ideal bilayer structure and $B_{11}N_{11}C_{72}$. The final relaxed structures are shown in Fig. 1(g) and 1(h), respectively. The minimum B−N($NH_3$) distances are 2.89 Å for $NH_3@B_{36}N_{36}C_{72}$ and 1.58 Å for $NH_3@B_{11}N_{11}C_{72}$. Considering these structural parameters and the adsorption energies of –0.78 eV and –2.29 eV (the bar graph in Fig. 3(a) and Tables VII and VIII), it can be inferred that $NH_3@B_{36}N_{36}C_{72}$ undergoes physical adsorption, while $NH_3@B_{11}N_{11}C_{72}$ exhibits chemical adsorption. Bader charge analysis for $NH_3@B_{36}N_{36}C_{72}$ (Table V) reveals only minimal charge transfer, consistent with physical adsorption, and the Fermi level remains near the Dirac point (Figure 2(g)). For $NH_3$ adsorption on the non-ideal $B_{11}N_{11}C_{72}$ structure, the transferred charge in the graphene plane is 0.0373e (the bar graph in Fig. 3(b, d) and Table VI). The Fermi level shift is negligible (Fig. 2(h)). In the $NH_3@B_{11}N_{11}C_{72}$ system, charge predominantly transfers from the B−N plane to the nitrogen atom of ammonia. Notably, during adsorption, both the adsorbed molecule and the graphene subsystem act as acceptors. These results suggest that graphene–BN heterostructures are less sensitive for ammonia detection. A comparison of our results for the $NH_3@B_{36}N_{36}C_{72}$ system with DFT calculations of ammonia adsorption on pristine graphene [43] reveals an analogy similar to that observed for $NO_2$. In both cases, the absolute value of the charge transferred upon $NH_3$ adsorption is considerably smaller, and the corresponding shift of the Fermi level is negligible, in sharp contrast to the pronounced charge transfer and substantial Fermi-level shift induced by $NO_2$ adsorption. At the same time, ammonia adsorption in system $NH_3@B_{11}N_{11}C_{72}$ exhibits acceptor-type behavior, which may appear unusual. Although ammonia is commonly regarded as an electron donor on graphene, some first-principles studies have shown that on pristine h-BN it can act as a weak electron acceptor [47], gaining a small amount of charge from the substrate. In the present work, the BN layer is neither ideal nor electronically isolated: structural relaxation induces local distortions, while the underlying graphene substrate provides an additional channel for charge redistribution and Fermi-level alignment. As a result, the BN layer becomes electronically polarized, which facilitates charge transfer toward the adsorbed $NH_3$ molecule. This interface-induced polarization provides an explanation for the acceptor-type behavior of ammonia observed in our calculations.

**Conclusion**

Density functional theory (DFT) calculations with periodic boundary conditions were employed to investigate the gas-sensing behavior of two graphene/hexagonal boron nitride (h-BN) heterostructures upon adsorption of $NO_2$, $NH_3$, and $O_3$ molecules. The calculations were performed using the VASP package with the projector augmented-wave (PAW) method and the PBEsol functional, including DFT-D3 dispersion corrections. The model supercells comprised a graphene layer containing 72 carbon atoms, while the h-BN overlayer consisted of either 36 B and 36 N atoms or 11 B and 11 N atoms. The results reveal substantial differences between the ideal bilayer structure and the $B_{11}N_{11}C_{72}$ complex. Adsorption of $NO_2$, $NH_3$, and $O_3$ on the ideal bilayer structure corresponds to physisorption, whereas on the $B_{11}N_{11}C_{72}$ surface, the interaction is of a chemisorptive nature. In the latter, dissociative adsorption of ozone occurs, yielding O and O–O charged complexes. These findings are supported by the notably higher adsorption energies for the $B_{11}N_{11}C_{72}$ system and by structural analyses of the $M@B_{36}N_{36}C_{72}$ and $M@B_{11}N_{11}C_{72}$ complexes. It should be noted that when the dangling bonds in the $B_{11}N_{11}C_{72}$ system (the BN planar fragment) are passivated by hydrogen atoms ($C_{72}B_{11}N_{11}H_{12}$), the structure no longer exhibits pronounced distortion. Consequently, both the adsorption energies and the charge-transfer values obtained in our calculations become close to those of an ideal h-BN plane. These results indicate that the charge transfer upon adsorption is primarily governed by the local BN environment in the immediate vicinity of the adsorbed molecule. The sensing capability of the investigated heterostructures can be assessed through the relative shift between the Fermi level and the Dirac point induced by gas adsorption. For $NO_2$ and $O_3$, the Fermi-level shift leads to an increase in the partial density of electronic states within the graphene subsystem, thereby enhancing its electrical conductivity. Charge-transfer analysis corroborates this trend. Conversely, $NH_3$ adsorption induces only minor electronic perturbations, with a charge transfer of approximately 0.03 e, nearly an order of magnitude smaller than that observed for $NO_2$ and $O_3$ molecules.

**CRediT authorship contribution statement**

**Martin Siebel**: Investigation, Formal analysis, **Pavel Rubin** - Methodology, Formal analysis, Writing – original draft, Supervision, **Raivo Jaaniso** - Conceptualization, Writing – review and editing, Funding acquisition.

**Declaration of competing interest**

The authors declare that they have no known competing financial interests or personal relationships that could have appeared to influence the work reported in this paper.

**Acknowledgments**

This work was supported by the Estonian Research Council grants PRG1580 and Tem-Ta110. The simulated structures were visualized using the VESTA program [48]. Electronic charges were calculated using the Bader Charge Analysis code [49]. We thank Dr. I. Renge for helpful comments and discussion.

**Figures**

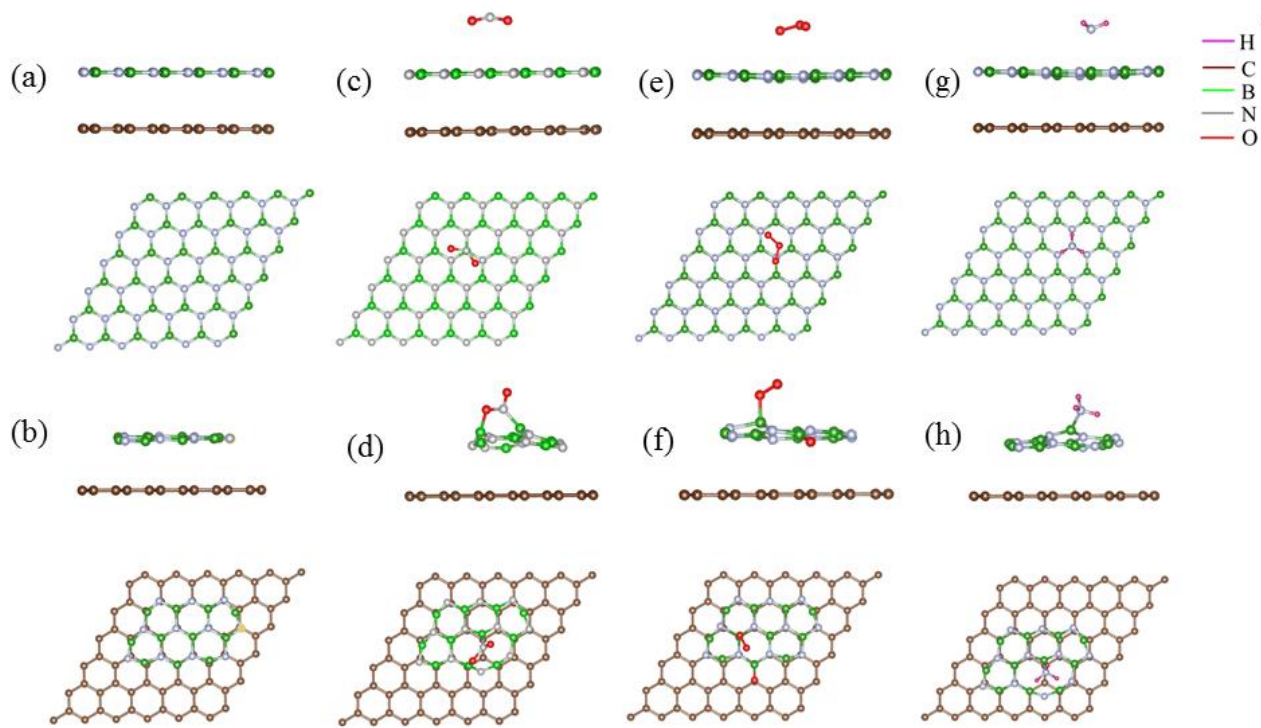

Fig. 1 A schematic view of the ground state geometric structures of the complexes $B_{36}N_{36}C_{72}$ (a), $B_{11}N_{11}C_{72}$ (b), $NO_2@B_{36}N_{36}C_{72}$ (c), $NO_2@B_{11}N_{11}C_{72}$ (d), $O_3@B_{36}N_{36}C_{72}$ (e), $O_3@B_{11}N_{11}C_{72}$ (f), $NH_3@B_{36}N_{36}C_{72}$ (g), $NH_3@B_{36}N_{36}C_{72}$ (h). **K**-mesh 4 x 4 x 1 mesh is used in calculations. Lattice parameters are 14.75, 14.75 and 30 Å.

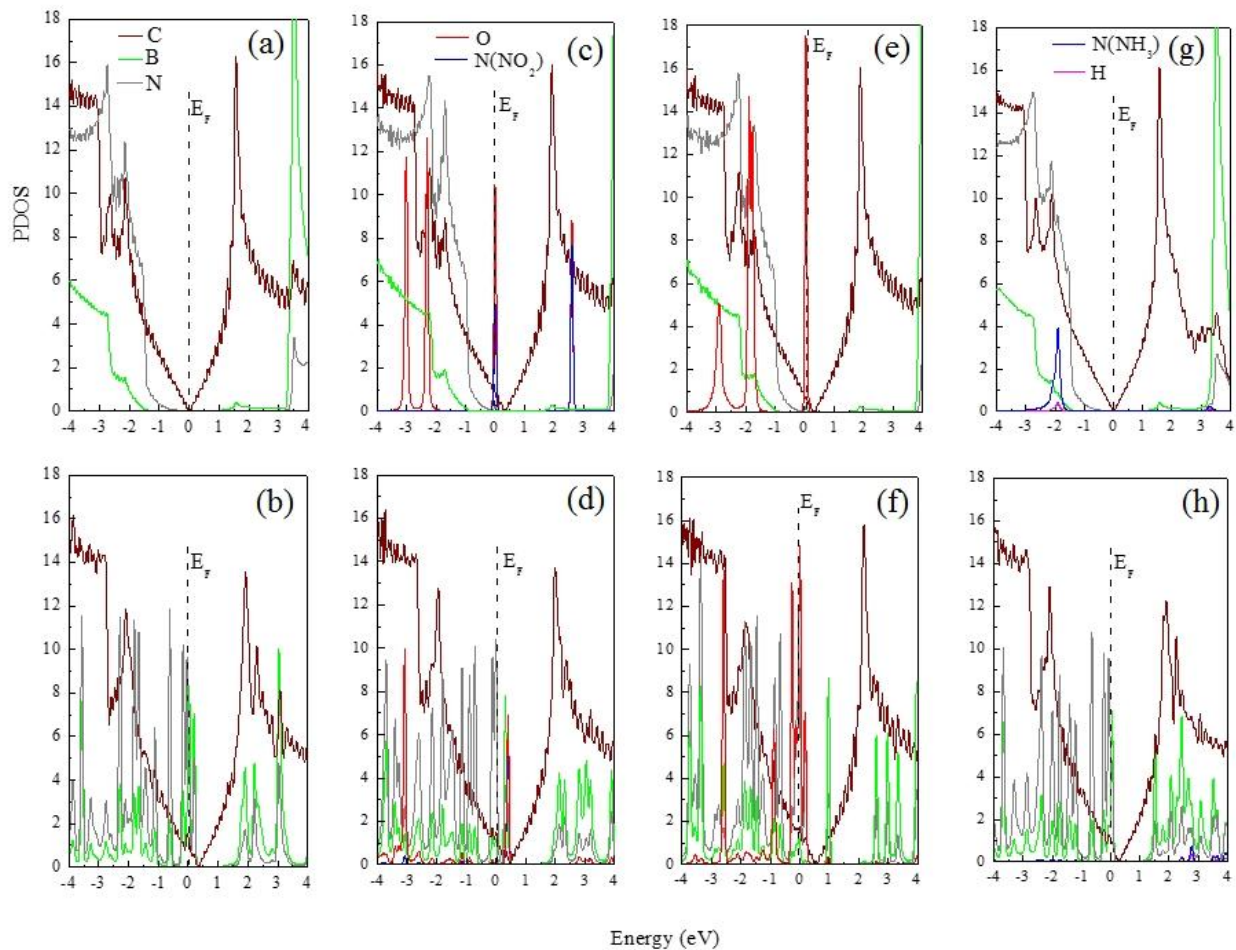

Fig. 2 Partial densities of the states of complexes $B_{36}N_{36}C_{72}$ (a), $B_{11}N_{11}C_{72}$ (b), $NO_2@B_{36}N_{36}C_{72}$ (c), $NO_2@B_{11}N_{11}C_{72}$ (d), $O_3@B_{36}N_{36}C_{72}$ (e), $O_3@B_{11}N_{11}C_{72}$ (f), $NH_3@B_{36}N_{36}C_{72}$ (g), $NH_3@B_{11}N_{11}C_{72}$ (h). All results were obtained using a **K**-mesh $20 \times 20 \times 1$. The solid lines at zero energy show the position of the Fermi level.

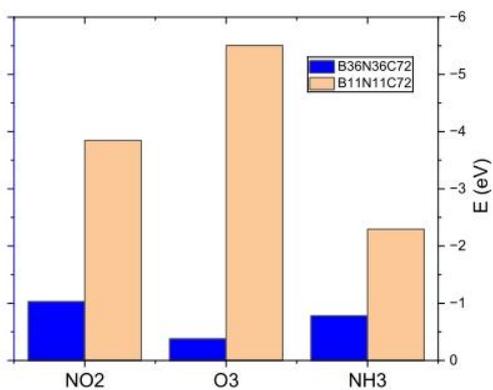
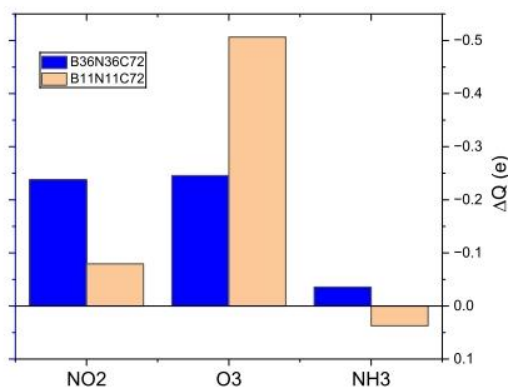
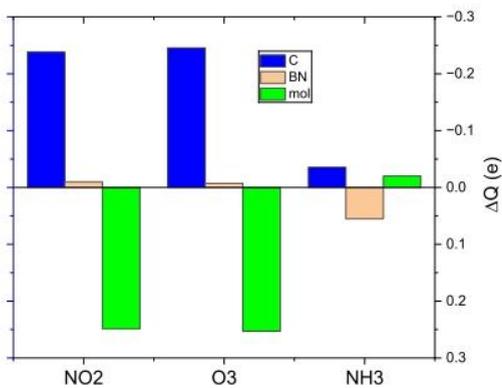
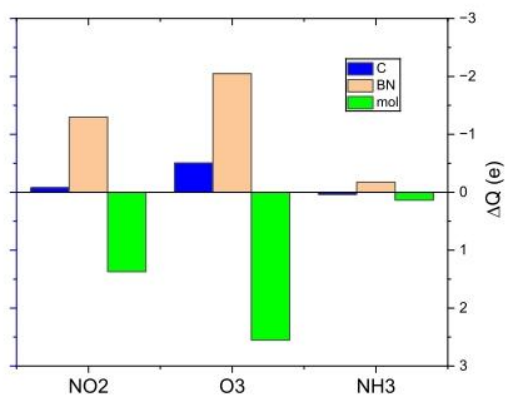

Fig. 3 Adsorption energies of $NO_2$, $O_3$, and $NH_3$ (a) and respective charge changes in the graphene layer (b) for two h-BN/Gr structures; charge changes in graphene, BN layer and adsorbed molecule for $B_{36}N_{36}C_{72}$ (c) and $B_{11}N_{11}C_{72}$ (d) structures.

*Supplementary Material*

# Gas sensing potential of stacked graphene/h-BN structures: a DFT-based investigation

## Martin Siebel, Pavel Rubin*, Raivo Jaaniso

Institute of Physics, University of Tartu, W. Ostwaldi 1, 50411 Tartu, Estonia

*rubin@ut.ee

Table I Charge distribution in the $NO_2@B_{36}N_{36}C_{72}$ system. Bader analysis.

| Element | $B_{36}N_{36}C_{72}$ | | Single molecule $NO_2$ | | $NO_2@B_{36}N_{36}C_{72}$ | | Charge transfer due to adsorption |
|---|---|---|---|---|---|---|---|
| | Initial | Bader charge analysis | Initial | Bader charge analysis | Initial | Bader charge analysis | |
| C | 288 | 288.0113 | | | 288 | 287.7733 | -0.238 |
| B | 108 | 0.0007 | | | 108 | 0.0004 | -0.0003 |
| N | 180 | 287.9879 | | | 180 | 287.9783 | -0.0096 |
| N(ads) | | | 5 | 4.205 | 5 | 4.2765 | 0.0715 |
| O | | | 12 | 12.794 | 12 | 12.9714 | 0.1774 |

Table II Charge distribution in the $NO_2@B_{11}N_{11}C_{72}$ system. Bader analysis.

| Element | $B_{11}N_{11}C_{72}$ | | Single molecule $NO_2$ | | $NO_2@B_{11}N_{11}C_{72}$ | | Charge transfer due to adsorption |
|---|---|---|---|---|---|---|---|
| | Initial | Bader charge analysis | Initial | Bader charge analysis | Initial | Bader charge analysis | |
| C | 288 | 287.7272 | | | 288 | 287.6478 | -0.0794 |
| B | 33 | 0.0005 | | | 33 | 0.000368 | -0.000132 |
| N | 55 | 88.27228 | | | 55 | 86.9756 | -1.2967 |
| N(ads) | | | 5 | 4.205 | 5 | 4.8958 | 0.6908 |
| O | | | 12 | 12.794 | 12 | 13.4804 | 0.6864 |

Table III Charge distribution in the $O_3@B_{36}N_{36}C_{72}$ system. Bader analysis.

| Element | $B_{36}N_{36}C_{72}$ | | Single molecule $O_3$ | | $O_3@B_{36}N_{36}C_{72}$ | | Charge transfer due to adsorption |
|---|---|---|---|---|---|---|---|
| | Initial | Bader charge analysis | Initial | Bader charge analysis | Initial | Bader charge analysis | |
| C | 288 | 288.0113 | | | 288 | 287.7657 | -0.2456 |
| B | 108 | 0.0007 | | | 108 | 0.00045 | -0.00025 |
| N | 180 | 287.9879 | | | 180 | 287.981 | -0.0069 |
| O | | | 18 | 18 | 18 | 18.2528 | 0.2528 |

Table IV Charge distribution in the $O_3@B_{11}N_{11}C_{72}$ system. Bader analysis.

| Element | $B_{11}N_{11}C_{72}$ | | Single molecule $O_3$ | | $O_3@B_{11}N_{11}C_{72}$ | | Charge transfer due to adsorption |
|---|---|---|---|---|---|---|---|
| | Initial | Bader charge analysis | Initial | Bader charge analysis | Initial | Bader charge analysis | |
| C | 288 | 287.7272 | | | 288 | 287.2208 | -0.5064 |
| B | 33 | 0.0005 | | | 33 | 0.000246 | -0.000254 |
| N | 55 | 88.27228 | | | 55 | 86.22453 | -2.04775 |
| O | | | 18 | 18 | 18 | 20.5544 | 2.5544 |

Table V Charge distribution in the $NH_3@B_{36}N_{36}C_{72}$ system. Bader analysis.

| Element | $B_{36}N_{36}C_{72}$ | | Single molecule $NH_3$ | | $NH_3@B_{36}N_{36}C_{72}$ | | Charge transfer due to adsorption |
|---|---|---|---|---|---|---|---|
| | Initial | Bader charge analysis | Initial | Bader charge analysis | Initial | Bader charge analysis | |
| C | 288 | 288.0113 | | | 288 | 287.9756 | -0.0357 |
| B | 108 | 0.0007 | | | 108 | 0 | -0.0007 |
| N | 180 | 287.9879 | | | 180 | 288.0439 | 0.056 |
| N(ads) | | | 5 | 7.9998 | 5 | 7.98 | -0.0198 |
| H | | | 3 | 0.0002 | 3 | 0.0005 | 0.0003 |

Table VI Charge distribution in the $NH_3@B_{11}N_{11}C_{72}$ system. Bader analysis.

| Element | $B_{11}N_{11}C_{72}$ | | Single molecule $NH_3$ | | $NH_3@B_{11}N_{11}C_{72}$ | | Charge transfer due to adsorption |
|---|---|---|---|---|---|---|---|
| | Initial | Bader charge analysis | Initial | Bader charge analysis | Initial | Bader charge analysis | |
| C | 288 | 287.7272 | | | 288 | 287.7645 | 0.0373 |
| B | 33 | 0.0005 | | | 33 | 0.000374 | -0.000126 |
| N | 55 | 88.27228 | | | 55 | 88.09808 | -0.1742 |
| N(ads) | | | 5 | 7.9998 | 5 | 8.13657 | 0.1368 |
| H | | | 3 | 0.0002 | 12 | 0.000438 | 0.000238 |

Table VII Adsorption energies of the $M@B_{36}N_{36}C_{72}$ systems.

| Adsorbed species | Single molecule | $B_{36}N_{36}C_{72}$ | $M@B_{36}N_{36}C_{72}$ | Adsorption energy |
|---|---|---|---|---|
| $NO_2$ | -18.410 | | -1382.2734 | -1.029 |
| $O_3$ | -14.4771 | -1362.8344 | -1377.6874 | -0.3759 |
| $NH_3$ | -18.947 | | -1382.564 | -0.7826 |

Table VIII Adsorption energies of the $M@B_{11}N_{11}C_{72}$ systems.

| Adsorbed species | Single molecule | $B_{11}N_{11}C_{72}$ | $M@B_{11}N_{11}C_{72}$ | Adsorption energy |
|---|---|---|---|---|
| $NO_2$ | -18.410 | | -898.2132 | -3.8469 |
| $O_3$ | -14.4771 | -875.9563 | -895.9384 | -5.5053 |
| $NH_3$ | -18.947 | | -897.198 | -2.295 |
15